# Kinetically-stabilized Ferroelectricity in Bulk Singlecrystalline HfO$_2$:Y without Wake-up Effects

*Xianghan Xu, Fei-Ting Huang, Yubo Qi, Sobhit Singh, Karin M. Rabe, Dimuthu Obeysekera, Junjie Yang, Ming-Wen Chu, and Sang-Wook Cheong*

Xianghan Xu, Fei-Ting Huang, Sang-Wook Cheong
Rutgers Center for Emergent Materials and Department of Physics and Astronomy,
Rutgers University, Piscataway, NJ 08854, USA
E-mail: sangc@physics.rutgers.edu

Yubo Qi, Sobhit Singh, Karin M. Rabe
Department of Physics and Astronomy,
Rutgers University, Piscataway, NJ 08854, USA

Dimuthu Obeysekera, Junjie Yang
Department of Physics,
New Jersey Institute of Technology, Newark, NJ 07102, USA

Ming-Wen Chu
Center for Condensed Matter Sciences and Center of Atomic Initiative for New Materials,
National Taiwan University, Taipei 106, Taiwan

## Abstract

HfO$_2$, a simple binary oxide, holds ultra-scalable ferroelectricity integrable into silicon technology. Polar orthorhombic (*Pbc*2$_1$) form in ultra-thin-films ascribes as the plausible root-cause of the astonishing ferroelectricity, which has thought not attainable in bulk crystals. Though, perplexities remain primarily due to the polymorphic nature and the characterization challenges at small-length scales. Herein, utilizing a state-of-the-art Laser-Diode-heated Floating

Zone technique, we report ferroelectricity in bulk singlecrystalline $HfO_2$:Y as well as the presence of anti-polar *Pbca* phase at different Y concentrations. Neutron diffraction and atomic imaging demonstrate (anti-)polar crystallographic signatures and abundant 90°/180° ferroelectric domains in addition to the switchable polarization with little wake-up effects. Density-functional theory calculations suggest that the Yttrium doping and rapid cooling are the key factors for the desired phase. Our observations provide new insights into the polymorphic nature and phase controlling of $HfO_2$, remove the upper size limit for ferroelectricity, and also pave a new road toward the next-generation ferroelectric devices.

## Introduction

As a quantum-leap for the miniaturization of semiconductor circuits, hafnium-based high-k dielectrics have been implemented as an alternative to thin silicon dioxide gate insulators in silicon technology since 2007[1-4]. Remarkably, the continued dimensional scaling-down lead to the observation of unexpected ferroelectricity (FE) in ultrathin $HfO_2$ films in both pure and doped forms[5-7]. The good CMOS-compatibility, as well as the robust ferroelectricity down to few layers[8], has strongly suggested hafnia a promising candidate for the next generation of non-volatile memory devices beyond the PZT-based ferroelectrics[9-11]. In parallel, the application of hafnia-based ferroelectrics as gate insulators into negative capacitance transistors offers a great potential for extending Moore's law[12-14].

In the light of the extensive research on this relatively simple binary oxide over the past few decades, the first observation of ferroelectric property in $HfO_2$:Si thin film at room temperature was totally unexpected[6]. The known six equilibrium phases of hafnia are all non-polar, and hence, incapable of supporting ferroelectricity[15]. These six phases include the room temperature monoclinic (*m*) $P2_1/c$ phase, the high temperature tetragonal (*t*) $P4_2/nmc$ and cubic (*c*) *Fm-3m* phases, and the high-pressure fluorite $CaF_2$-type anti-polar orthorhombic *Pbca* (*o*-AP or *o*-I), disordered *Pbcm* (*o*-DO) phases, and the cotunnite ($PbCl_2$)-type *Pmnb* (*o*-II) phase. The results of prior computational and empirical investigations hinted at the non-equilibrium polar orthorhombic $Pca2_1$ (*o*-FE) phase as the plausible root-cause of the astonishing ferroelectricity in hafnia[16]; however, different phases, in addition to many unexplored metastable phases, have also been experimentally identified[17-19]. Perplexities and confusions arise primarily due to the polymorphic nature of hafnia and also due to the challenges associated with the characterization of the mixed/complex phases in ultrathin films at small length scales[20]. The different sets of growth conditions, including dopant species[5,6,21-23], substrates[17,24-26], and capping electrodes[6,21,27] all add some degree of uncertainty in these characterizations. Furthermore, various extrinsic factors such as stress[28], oxygen vacancies[29,30], surface energy[31], and electric fields[32] have been put forward to provide a rationale for the ferroelectricity[33,34].

In any case, a small-size-driven tetragonal to ferroelectric phase transition at the nanoscale has been suggested to be essential to stabilize the $o$-FE phase in ultra-thin films[21,35,36], which is unlikely to happen in bulk crystals. Even though the $o$-FE phase was first reported in bulk polycrystalline Mg-doped $ZrO_2$ at 30 K[37], the ferroelectric phase in bulk hafnia (zirconia) was presumed not to be stable at ambient pressure at room temperature[31,34,38]. Therefore, for the further exploitation of hafnia-based science and technologies, it is quintessential to understand the (meta)stability of the ferroelectric phase in hafnia with doping under various environments in its "bulk forms". Emphasize that in spite of world-wide efforts on the scientific and technological investigations on (doped) hafnia, the ferroelectricity in bulk (doped) hafnia has never been reported. Hafnia is extremely refractory with a melting temperature of 2,758 °C, and this high thermal stability is a part of the reasons for the limited recent research on bulk (doped) hafnia. Herein, we demonstrate ferroelectricity in bulk $HfO_2$:Y single crystals, which are grown utilizing a state-of-the-art laser-diode-heated floating zone (LDFZ) technique. A comprehensive study on the structural phase transition and microstructures of those orthorhombic polymorphs using aberration-corrected scanning transmission electron microscopy (STEM) and dark-field (DF) TEM technique demonstrates the presence of abundant microstructures including $90^o$ and $180^o$ ferroelectric domains. In addition, we are able to stabilize another orthorhombic $o$-AP phase, which is known as the high-pressure orthorhombic anti-polar phase, at room temperature and ambient pressure. Surprisingly, the ferroelectricity in our bulk crystals does not accompany any "wake-up effect", while it is commonly observed in ultra-thin films. Our results provide an essential stepping-stone toward the scientific understanding of the dielectric multiplicity of (doped) hafnia toward future hafnia-based science and technologies.

# Main

**Growth and structural characterization of bulk HfO$_2$:Y**

To stabilize the metastable phases in bulk, whose appearance cannot be anticipated from the traditional phase diagrams, the non-equilibrium growth through a quenching process is the key[39]. Laser-Diode-heated Floating Zone furnace (LDFZ) having the merit of well-focused laser diodes as heating sources with maximum temperature around 3,000°C, i.e., a steep temperature gradient at the growth interface enables a much higher growth and quenching speed than the conventional Infrared-Lamp FZ technique[40]. For the non-equilibrium growth, a fiber feed rod (3 mm) was adjusted and pulled at a relatively high speed 20 mm/h from the melt (see Methods). Figure 1a shows the lab X-ray diffraction pattern (XRD) for our HfO$_2$:Y crystals with different Yttrium doping concentrations. As revealed, the ground state of pure HfO$_2$ crystal is monoclinic, Y-rich ends in the cubic phase, similar to the famous Yttrium-stabilized zirconia (YSZ)[41]. Surprisingly, new peaks appear (at 2θ = 35°, 43.5°, and 75°) within the 8-12 % Y doping while they disappear after a high-temperature (1,600°C) annealing, indicating the metastability of the new phases (Supplementary Figure 1). 12 % HfO$_2$:Y is particularly interesting as the monoclinic phase is fully suppressed, accompanying the disappearance of $(11\text{-}1)_m$ and $(111)_m$ reflections at 2θ = 28.5° and 31.7° (Figure 1a). The profile fitting leads to tetragonal/orthorhombic unit cells with aspect ratio $c/a$ or $2b/(a+c) \approx 1.01$ and possible candidates of either $t$, $o$-AP, $o$-DO, or $o$-FE phases. The similarity comes that $o$-AP can be considered alternate stacks of $o$-FE domains and resulting in the $a$-lattice doubling while the $o$-DO can be treated as a macroscopic average of $o$-FE domains[42].

A representative crystal image of 12% HfO$_2$:Y is presented in Figure 1b. Up to 50 mm long HfO$_2$:Y transparent crystal boule can be grown. Transmission polarized optical microscope (tPOM shown in Figure 1c) reveals a tweed texture on a cutting surface, namely a crisscross of needle-like domains, which is typical in orthorhombic systems with fine orthorhombic twins. The twin character can be recognized in the fact that domain contrast reversal by manipulating the analyzer under tPOM. We notice 8-12 % HfO$_2$:Y crystals share some general features of tweed textures but whose details can vary (Supplementary Figure 2). The orthogonal nature of the tweed textures is further confirmed by selected area electron diffraction (SAED) patterns under TEM (Figures 1e-1g) which rule out the presence of the $t$ phase. The cell-doubling superlattice peaks, i.e. *o*-AP, persist in 8-11 % range which can be clearly identified through a sequence of SAED patterns (Figure 1e and Supplementary Figure 3), whereas they disappear in the 12 % $HfO_2$:Y crystal (Figure 1f and Supplementary Figure 4). A closer examination of those SAED patterns in the 12 % $HfO_2$:Y crystal reveals no peak splitting or diffuse streaking attributing to the formation of mixed phases or defects. Though the same extinction rules in polymorph *o*-DO and *o*-FE yield ambiguity to distinguish them in 12 % $HfO_2$:Y using three-dimensional SAEDs; instead, neutron scattering is the solution and notably, it is not applicable for thin-film investigations. The result is discussed in detail in the following section. The comprehensive analysis of structural and microstructure properties leads to the non-equilibrium phase diagram (Figure 1d), which is distinct from the known $HfO_2$-$Y_2O_3$ binary phase diagram reflecting equilibrium states[43]. Thus, the polycrystalline ceramics and the LDFZ single crystals can have quite different phases even they have the same Y doping concentration (Supplementary Figure 1), and those intriguing metastable orthorhombic phases can be stabilized, for the first time, as we demonstrate in this work.

### *o*-FE neutron refinement

Next, the *o*-FE phase of 12 % $HfO_2$:Y is determined by a series of single or multiphase Rietveld refinements of neutron powder diffraction (NPD) patterns. The use of neutron which penetrates massive crushed powder to depths of centimeter ensures the precision and representativeness of the oxygen distribution. For the NPD measurements, approximately a 2-gram single crystal was finely ground and loaded in a vanadium can of 3 mm diameter. NPD measurements were performed on the time-of-flight powder diffractometer, POWGEN, located at the Spallation Neutron Source at Oak Ridge National Laboratory. The data were collected with neutrons of central wavelength 0.8 Å at room temperature. The NPD patterns were refined using the General Structure Analysis System (GSAS)[44,45]. Lattice parameters are derived by refinement, displayed in Table 1. The observed diffraction intensities in Figure 2a is fitted by 100 % *o*-FE orthorhombic phase with $\chi^2$=15.08 % and lattice parameter a=5.1019 Å, b=5.154 Å, c=5.1036 Å. For the sake of simplicity and comparison with *o*-DO (*Pbcm*)/*o*-AP (*Pbca*), we adopt $Pbc2_1$ in which lattice *a* and *b* are switched related to $Pca2_1$ and the net polar displacement remains along the *c*-axis. The concise main peak fitting in Figure 2b suggests a reliable refinement. To further verify that *o*-FE is the major phase, we carefully compared the calculated intensities with possible coexisting phases with analogous structures such as *o*-DO and *t* phase. First, the existence of peaks around 2.75 Å$^{-1}$ and 3.68 Å$^{-1}$ which cannot be explained using the *t* phase (blue profile in Figure 2c) indicates a lower symmetry than the *t* phase. The significant intensity of the peak around 3 Å$^{-1}$ denoted in Figure 2d can rule out the *o*-DO phase. Thus, the neutron diffraction data confirm the polar *o*-FE phase in 12 % HfO$_2$:Y crystals.

**Ferroelectricity and 90°/180° ferroelectric domains**

The bulk orthorhombicity which defined by $\epsilon = (c-a)/(c+a)*100\% = 0.017\%$ in our work is about six times smaller than the reported value 0.099 % in the HfO$_2$:Gd thin film[16]. The small orthorhombicity explains in part the tweed textures observed in tPOM (Figure 1c) and DF-TEM image along [010]$_{o\text{-FE}}$ (Supplementary Figure 5). Those fine *ac* twins occur because of the ferroelastic spontaneous strain and are characterized by a crosshatching up to a few hundred nanometers in length. The atomically resolved positions of the projected Hf atom sublattice and the twin boundary are readily determined using high-angle annular dark-field (HAADF) STEM which displays strong contrast associated with the heavy Hf atoms (Figure 3a-3b). Figures 3c-3d indicate a 90° rotation of the diffractograms by fast Fourier transform (FFT) operations of green and red selected areas in Figure 3a. In real space, the periodic short-long Hf-Hf bonding along the *a*-axis revealed by overlaid color-coded *o*-FE unit cells also rotates 90° across the twin boundary oriented along [101]$_{o\text{-FE}}$. Since the polarization is along the *c*-axis, these *ac* twin walls are 90°-type FE domain walls. Consequently, the microstructures perpendicular to [010]$_{o\text{-FE}}$, i.e. [100]$_{o\text{-FE}}$ and [001]$_{o\text{-FE}}$ are similar, with abundant irregular shaped *ab* and *bc* twin domains within a few hundred nanometers in size (Supplementary Figure 5). There is almost an equal distribution of *ab* and *bc* domains. A further zoomed-in HAADF-STEM image of one *ab*/*bc* twin wall (Figure 3b) shows an atomically coherent wall along the longest *b*-axis. The crystallographic orientation of *ab*/*bc* domains is confirmed by the diffractograms of purple and blue selected areas (Figures 3e-3f) by FFT operations. Inside those ferroelastic *ac*/*ab*/*bc* domains, there exist 180°-type FE domains in a hundred nanometer scale or less. Figures 3g-3h display the DF-TEM images of FE domains with bright and grey contrasts viewing along [111]$_{o\text{-FE}}$ direction, resulting from the non-equal diffraction intensity due to the broken space-inversion of the o-FE phase. The domains with two different contrasts are associated with the ±c polar axes as indicated by white arrows. The yellow arrows mark the ferroelastic domain walls, in which the contrast remains deep-dark when ±g = (11$\bar{2}$) spots were excited (Figures 3g-3h). Our DF-TEM images demonstrate consistently the presence of 180° and 90° FE domains.

Figure 3i shows polarization *vs* electric field P-E hysteresis loop of 12 % $HfO_2$:Y crystal on such a [100]/[001]$_{o\text{-FE}}$ oriented thin piece as shown in Figure 3b, showing a switchable polarization at room temperature. The net remnant polarization $2P_r$ is estimated to be 4 μC/cm$^2$ with $E_c = 2$ MV/cm. The P-E loops measured with different frequencies show same hysteresis behaviors (Supplementary Figure 6), undoubtedly confirming they really arise from ferroelectricity. Field cycling test is performed by applying poling pulses on a [100]/[001]$_{o\text{-FE}}$ piece of 12 % $HfO_2$:Y crystal. For each data point, total polarization $P_{total}$, non-remnant polarization $P_{non\text{-}remnant}$, and remnant polarization $P_r$ were measured by PUND method. As Figure 3j shows, no significant enhancement of $P_r$ is observed under up to $10^7$ cycling. Note that the increase of $P_{total}$ and $P_{non\text{-}remnant}$ indicates that the sample becomes dielectric lossy during the test, while the ferroelectric component $P_r$ remains constant. Therefore, the 12 % $HfO_2$:Y single crystal doesn't exhibit any wake-up effect, which refers to increasing remnant polarization with increasing field cycling.

**Theoretical $P_r$ value and no wake-up effect**

The measured value of the polarization is much smaller than the theoretical value for a pristine bulk *o*-FE $HfO_2$ (61 μC/cm$^2$ given by the DFT calculation). This is consistent with the experimentally determined structure (Table 1), as the polarization calculated for this structure from DFT is only 9 μC/cm$^2$, which matches fairly well with the experimental result (4 μC/cm$^2$). The suppression of polar distortion can be understood as resulting from Y substitution with associated charge-compensating oxygen vacancies. In our DFT calculations, we introduce 12.5 % Y doping and 3.25 % oxygen vacancy into the pristine *o*-FE $HfO_2$ and then relax the structure. The neutron diffraction simulation given by the optimized structure matches with the experimental one very well (Supplementary Figure 7), indicating the robustness of the DFT method. We find that the optimized structure exhibits a polar distortion $Q_P$ of only 0.015 Å (See Supplementary Figure 8 for the definition of $Q_P$), which is in the same order of magnitude of the experimental result 0.035 Å (the corresponding comparison for undoped $o$-FE HfO$_2$ is 0.235 Å). Besides, the polarization values in Y doped HfO$_2$ thin films reported in other studies are also small (~5 μC/cm$^2$), which further supports our point[5,46].

Our sample exhibits as-grown ferroelectricity without any wake-up cycling. This is because our bulk HfO$_2$ is a single crystal in the $o$-FE phase with approximately uniformly distributed oxygen vacancy. Generally, the wake-up effect originates from two mechanisms: (1) the elimination of internal bias induced by the redistribution of oxygen vacancies[47,48] and (2) the non-polar to a polar phase transition[49,50]. However, neither of them applies to our case. Our bulk HfO$_2$ does not have interfaces where oxygen vacancies concentrate. At 12.5 % Y, the oxygen vacancies are distributed in association with the substituted Y, and the fast quenching also tends to prohibit the diffusion of oxygen vacancies. As a result, our sample is unlikely to possess any particular region pinned by a high oxygen vacancy concentration. Moreover, our bulk sample adopts an as-grown ferroelectric $o$-FE phase and does not have any electric-field-induced non-polar to polar phase transition. The absence of wake-up effect indicates clean and intrinsic bulk ferroelectric properties.

**Discussion about growth and kinetic theory**

In contrary to the bulk ferroelectricity demonstrated in this work, it has been reported that, in HfO$_2$ based thin films, the ferroelectricity keeps degrading and finally vanishes with increasing thickness, as a result of decreasing orthorhombic phase portion. The bottleneck to stabilize the orthorhombic phase in thicker films includes relaxation of strain from the substrate, and diffusion of dopant atoms causing phase separations. Our bulk crystal growth technique provides an alternative to overcome this bottleneck, by manipulating the thermal history. During growth, HfO$_2$:Y crystallizes from a molten liquid at temperatures above 2,700$^o$C, which produces a solid solution with uniform Yttrium distribution. Then, by a subsequent rapid cooling, the Yttrium atom diffusion is frozen resulting in an Yttrium supersaturation state. We believe the huge chemical pressure from Yttrium supersaturation drives the formation of metastable orthorhombic phases. 1,600$^o$C annealing on as-grown HfO$_2$:Y crystals results in the decomposition of the orthorhombic phase into cubic and monoclinic phases (Supplementary Figure 1), probably because the Yttrium ionic diffusion is activated again at such a high temperature. Annealing with moderate temperatures such as 1,000°C doesn't change the phase.

Specifically, our DFT calculations demonstrate that the bulk $o$-FE HfO$_2$ phase can be kinetically stabilized by the joint effect of quenching and Yttrium supersaturation. In Figure 4a, we plot the energy landscape of bulk HfO$_2$ as a function of the amplitudes of several characteristic modes (See the Supplementary Figure 8 for the definitions and detailed descriptions of these modes). The energy landscape contains several local minima, which corresponds to the stable or metastable $t$, $m$, $o$-AP, and $o$-FE phases. Here, instead of the $c$ phase, the $t$ phase is viewed as the high-temperature phase, since the $c$ phase is unstable in DFT calculation ($T = 0$ K) and automatically collapses to the $t$ phase[18,51]. This assumption is reasonable since the $t$ and $c$ phases share quite similar physical properties[17,50]. Note that the $o$-FE to cubic phase transition is observed around 800°C determined by in-situ heating TEM (Supplementary Figure 9). The clear thermal hysteresis during warming and cooling cycles is the characteristic of a first-order phase transition. No hint of tetragonal phase is detected experimentally.

The energy profiles of several representative reaction paths linking the high-temperature $t$ phase and three different low temperature phases ($m$, $o$-AP and $o$-FE) are shown in Figure 4b. Our DFT calculations reveal that even though the $o$-FE phase has the highest energy among the low-temperature phases, the $t$ to $o$-FE phase transition has the smallest energy barrier. Fast cooling (quenching) makes the reaction path with a smaller energy barrier more favorable. Moreover, supersaturated Y doping decreases the crystallinity, reduces the crystallite size, and increases the ratio between the interface/grain boundary area and volume of the nuclei of new phases[52]. To reduce the strain energy, the nuclei tend to have similar lattice constants as the parent tetragonal phase. We carry out DFT calculations to obtain the energy barriers of the representative paths shown in Figure 4, but with the fixed lattice constants as in the $t$ phase. Our results, listed in Table 2, clearly show that the energy barriers are enlarged and $t$ to $o$-FE is further favored when lattice constants are fixed. This implies that supersaturated Y doping can facilitate the formation of the $o$-FE phase. In addition, we notice that the $o$-AP always has the lower energy than $o$-FE no matter lattice constants fixed or not, indicating the $o$-AP thermodynamically more stable phase. This is consistent with the fact that $o$-AP exists widely (8 % to 11 %) in the HfO$_2$:Y crystal phase diagram (Figure 1d) whereas $o$-FE exists only in a narrow regime near 12%.

## Outlook

A series of HfO$_2$:Y single crystals are grown by LDFZ technique. Metastable orthorhombic *o*-FE (*Pbc*2$_1$) and *o*-AP (*Pbca*) phases are, for the first time, stabilized in bulk single crystals at room temperature and under ambient pressure. The relatively rapid cooling using our unique LDFZ technique is essential to stabilize the orthorhombic phase with switchable and scale-free polarization without substrates or strain. Neutron diffraction on 12 % HfO$_2$:Y suggests 100 %-pure polar *Pbc*2$_1$ phase. The switchable polarization is characterized by P-E loops showing 2$P_r$ = 4 μC/cm$^2$ and $E_c$ = 2 MV/cm, which is consistent with the theoretical value based on the experimentally determined crystallographic structure. Our DFT calculations also support a kinetically stabilized route toward the bulk *o*-FE HfO$_2$ phase through the combination of quenching, Yttrium doping and oxygen vacancies. Furthermore, TEM studies unveil 90° ferroelectric twin domains and 180° ferroelectric domains inside those twins, and identify the *o*-FE to cubic ferroelectric Curie transition. A recent work suggests that the *o*-AP plays an important role as the intermediate phase in the *o*-FE polarization switching process[53]. Therefore, a deep understanding on the *o*-AP phase will be the key to unravel the scale-free ferroelectricity mystery of HfO$_2$. We emphasize that the *o*-AP phase at room temperature and under ambient pressure is experimentally reported, also for the first time, in this work. Our finding removes the upper length-scale limit for ferroelectric HfO$_2$, unveils the abundant microstructures, and provides insights toward revealing alternative routes for polymorphic phase controlling of HfO$_2$.

## Method

**Sample growth**: The HfO$_2$:Y single crystals were grown using a Laser-Diode-heated Floating Zone furnace. Stoichiometric and high-purity powders of HfO$_2$ and Y$_2$O$_3$ (900°C overnight baked) were mixed in a mortar, pelletized and sintered at 1,500°C in a box furnace for 20 hours with one intermediate grinding. The sintered pellet was pulverized, poured into a rubber tube, and pressed into a rod shape under 8,000 PSI hydrostatic pressure. The compressed rod was sintered at 1,600°C in a box furnace for 10 hours. The crystals were grown at the speed of 20 mm/hour in 0.5 L/min flowing $O_2$. This growth speed is fast enough to prevent phase separation and slow enough to guarantee good crystallinity.

**P-E and wake-up measurements**: The P-E loop in Figure 3i was measured on a $[100]/[001]_{o\text{-FE}}$ polished 12 % $HfO_2$:Y crystal with thickness 5 μm by the "PUND" hysteresis loop method provided in the Ferroelectric Material Test System (RADIANT TECHNOLOGIES INC.). The measurement period was 4 ms (f = 250 Hz). The remnant-only polarization hysteresis loop is finally derived.

Field cycling test in Figure 3j was performed by applying poling pulses on a $[100]/[001]_{o\text{-FE}}$ polished 12 % $HfO_2$:Y crystal with thickness 5 μm. The poling pulses had voltage 1200 V, frequency 5000 Hz, and width 0.015 ms. The $P_{total}$, $P_{non\text{-}remnant}$, and $P_r$ for each data point were measured by the "PUND" method provided in the Ferroelectric Material Test System (RADIANT TECHNOLOGIES INC.) with switched pulse (opposite preset pulse), non-switched pulse (same preset pulse), and $P_r = P_{total} - P_{non\text{-}remnant}$, respectively.

**TEM measurements**: Specimens for (S)TEM studies were fabricated in four cutting single crystals including [111], [010], [100], and [001] axes by mechanical polishing, followed by Ar-ion milling and studied using JEOL-2010F TEM and JEOL-2000FX TEM. Due to the high-density $ac$-twins, [100] and [001] zone axes are usually entangled so that we probe the microstructures along three directions: [111], [010], and [100]/[001]. We demonstrate the existence of $ac$ twins, $ab$-and $bc$-domains of 12 % $HfO_2$:Y using DF-TEM and HAADF-STEM. HAADF-STEM imaging was conducted using a spherical aberration-corrected transmission electron microscope (JEOL-2100F) and acquired in two conditions: 512 ×512 with 0.019 nm and 0.015 nm/pixel with collection angle between 80–210 mrad. Though the average sizes of $ac$ twins, $ab$- and $bc$-domains are about a few hundred nanometers square, larger $ac$ twin domains still exist. To unveil the FE domains, we have prepared a few [111] cutting pieces and observed $o$-FE domains by DF-TEM imaging taking $g\pm = \pm 11\bar{2}$ along [111] zone axis with the polarization along the $\pm c$ directions. In-situ heating TEM experiment was carried out using a JEOL-2000FX TEM with a high-temperature specimen holder. All images are raw data.

**DFT calculations**: Density functional theory (DFT) calculations are carried out with the ABINIT and Quantum-espresso packages. The exchange-correlation functional is approximated with the local density approximation (LDA). The norm-conserving pseudopotentials are generated by the Opium package. The Brillouin zone is sampled with a 4 × 4 × 4 Monkhorst-Pack *k*-point mesh and the plane-wave cutoff energy is 50 Ry. The atomic force convergence threshold is set as 5× $10^{-6}$ Hartree per Bohr.

The energy landscapes are generated as functions of the lattice-mode amplitudes $Q_{A1}, Q_{A2},$ and $Q_P$ (See the Supplementary Figure 8 for the descriptions of these modes). For constructing the right part of the energy landscape, which is a function of $Q_{A2}$ and $Q_P$, we use a unit cell containing 12 atoms (4 unit formulas). We generate different structures with various amplitudes of the lattice-modes and then relax them with fixed mode amplitudes using the ABINIT package. Next, we fit the energies with the Landau type polynomials with the lattice-mode amplitudes as order parameters and use the polynomials to generate the 3-dimensional energy landscape. We follow the same procedures for generating the left part, but the unit cell contains 24 atoms (8 unit formulas) since the primitive cell of the *o*-AP phase is doubled compare with that of the *t*, *o*-FE, or *m* phase. The energy profiles of the representative reaction paths shown in Figure 4b are extracted from the energy landscapes in Figure 4a. The energy profiles (Supplementary Figure 10) for the systems with fixed lattice parameters are calculated with the nudged elastic band (NEB) method implemented in the Quantum-espresso package.

For the Y doped case, we substitute 2 Hf atoms with Y atoms in a 2 × 2 × 1 supercell, which corresponds to a 12.5 % Y doping. 2 Y dopants should be associated with 1 oxygen vacancy, in order to keep the valence balanced. We consider 32 different structures with the oxygen vacancy locating at the 32 possible sites, and optimize them with the DFT calculation. Finally, we select the one with the lowest energy as the representative structure. Due to the oxygen vacancy, the polar displacements of 4 Hf (or Y) atoms can not be calculated, since one of the eight oxygen atoms associated with a Hf atom is missing. Therefore, the polar distortion of such an Y doped structure is defined as the average of the 12 polar displacements corresponding to the Hf (or Y) atoms whose all eight oxygen neighbors remain.

## Acknowledgements

The experimental work was performed at the center for Quantum Materials Synthesis (cQMS), funded by the Gordon and Betty Moore Foundation's EPiQS initiative through grant GBMF6402, and by Rutgers University. Neutron diffraction studies used resources at the Spallation Neutron Source, a DOE Office of Science User Facility operated by the ORNL. The authors thank Dr. Qiang Zhang, from the ORNL, for support with neutron diffraction measurements. Y.Q., S.S. and K.M.R. were supported by the Office of Naval Research Grant N00014-17-1-2770. The DFT calculations performed for this study used the resources provided by the High-Performance Computing Modernization Office of the Department of Defense and the Rutgers University Parallel Computing (RUPC) clusters.

## Author contributions

X.X. and S.W.C. conceived the idea. X.X. synthesized the single crystals, collected the XRD data, and tested the ferroelectric properties. F.T.H. took the optical microscope pictures and conducted the TEM experiments. Y.Q., S.S., and K.R. performed the DFT calculations. D.O. and J.Y. carried out the neutron diffraction experiments and analysed the data. F.T.H. and M.W.C performed the STEM observations. X.X., F.T.H., Y.Q., S.S., K.R. and S.W.C. wrote the paper.

## Competing interests

The authors declare no competing interests.

## Additional information

Supplementary information is available for this paper

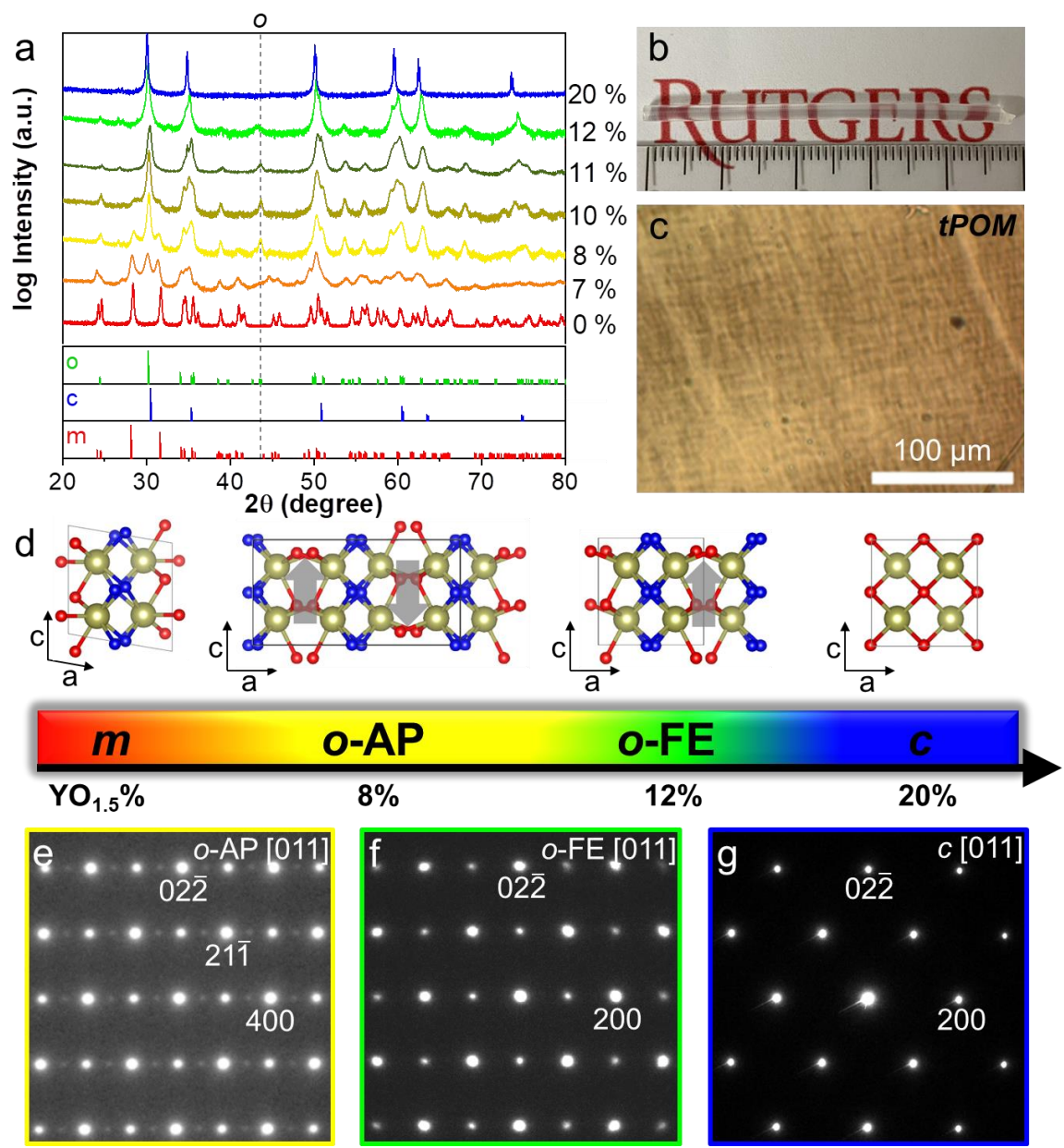

**Figure 1. Single crystals and phase diagram of HfO$_2$:Y. a**, Powder XRD on crushed crystals with various Y concentrations. The dashed line at 2θ = 43.5° denotes a peak uniquely belongs to *o* phase, without overlapping with peaks of *m* or *c*. **b**, Photograph of as-grown 12 % HfO$_2$:Y single crystal. **c**, Transmission polarized optical microscope (tPOM) image of one [010]-oriented 12 % HfO$_2$:Y polished crystal, showing fine *ac* twin features. **d**, A conceptual phase diagram of HfO$_2$:Y single crystals. The top panel shows the *ac*-projected structures of corresponding polymorphic phases. The reduced symmetry from the cubic phase lead to the displacements of Hf and two oxygen sites in *m*, *o*-AP, *o*-FE phases (oliver: Hf, red/blue: O). Grey arrows denote

the local polarizations in anti-polar (*o*-AP) and polar (*o*-FE) phases formed by the Hf/O displacements. **e-g,** Selected area electron diffraction patterns of 8 %, 12 % and 20 % HfO$_2$:Y samples along the [011] direction.

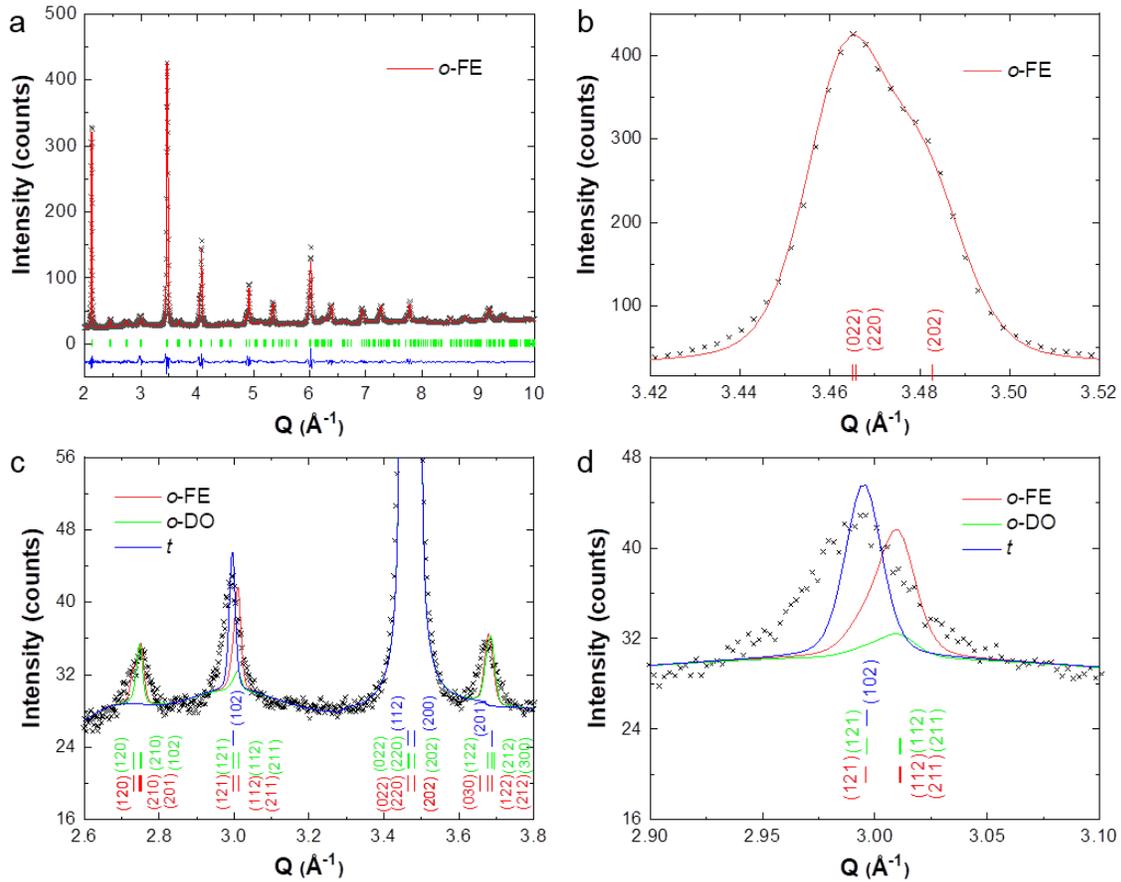

**Figure 2. Neutron diffraction studies report polar phase of 12 % HfO$_2$:Y single crystal. a,** Representative neutron diffraction pattern plotted against momentum transfer Q (Å$^{-1}$) = $2\pi/d$ at room temperature. The red profile through the black data points is a fit to the *o*-FE (*Pbc2$_1$*) structure. Green vertical bars indicate Bragg peak positions and the bottom blue curve shows the difference between the fit and the data. **b,** Detail of the {220} peaks, the peaks with the strongest intensity, showing a nice fit with the *o*-FE structure. **c,** Zoomed-in region showing a number of the weak reflections disallowed in the tetragonal *t* (blue curve: *P4$_2$/nmc*) and with lower calculated intensities in the disordered *o*-DO (green curve: *Pbcm*). Those peaks could be well described by *o*-FE (red curve). **d**, Zoomed-in region of **c** along with refined models and *hkl* indices below.

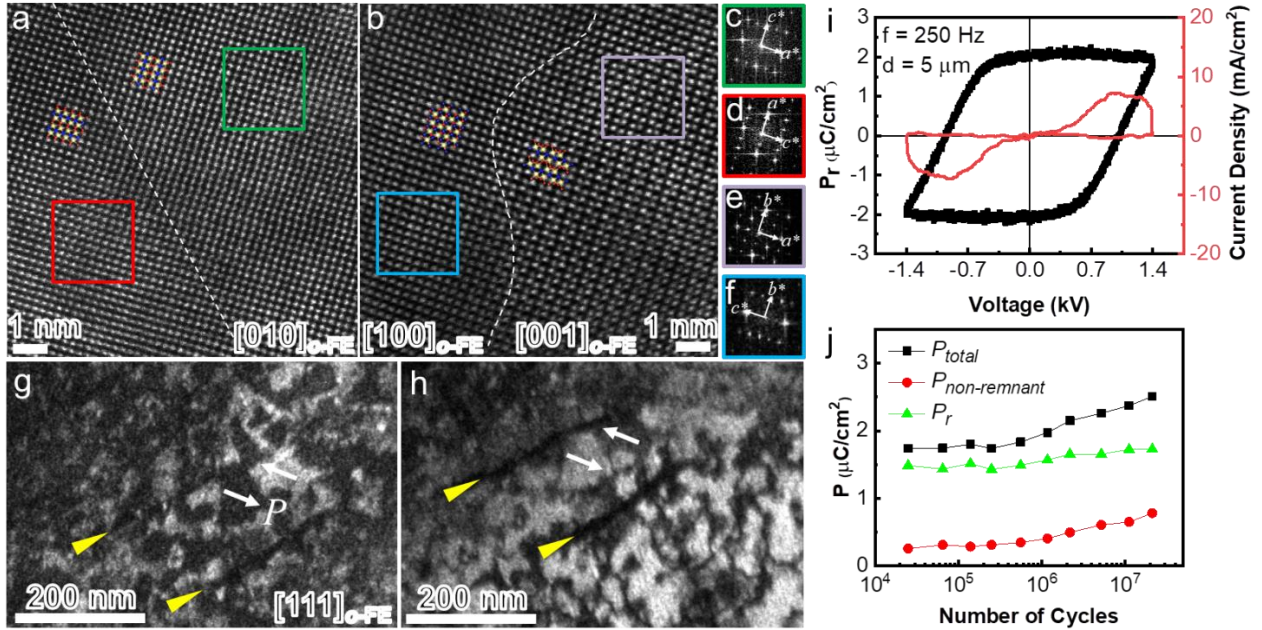

**Figure 3. Ferroelectric domains and electric polarization switching in 12 % $HfO_2$:Y. a**, High-angle annular dark-field STEM image along $[010]_{o\text{-FE}}$, showing *ac* twin domains and the straight domain wall. **b**, HAADF-STEM image along $[100]/[001]_{o\text{-FE}}$, showing irregularly shaped *ab* and *bc* twin domains and the curved domain wall. Domain walls indicated with dashed lines are added as a guide to the eye. *o*-FE models are superimposed (Hf: yellow; O: blue and red). **c-f**, The FFT diffractograms obtained from the selected areas of (a-b). **g**, Dark-field TEM image taken using the $g^+ = 11\bar{2}$ spot along $[111]_{o\text{-FE}}$. Abundant 180° FE domains at one-hundred-nm-scale or less are shown in black and white contrast due to the Friedel's pair breaking in *o*-FE. Black/white contrasts are associated with the $\pm c$ polar axes. **h**, The corresponding DF-TEM image taken using the $g^- = \bar{1}\,\bar{1}2$ spot with reverse contrast of 180° FE domains. Yellow arrows denote orthorhombic *ac* twin boundaries, in which contrast remains dark. **i**, The remnant P-E loop (black) and current density curve (red) measured by PUND method. **j**, The field cycling test with polarizations measured by PUND method. The black, red and green curves represent total polarization $P_{total}$, non-remnant polarization $P_{non\text{-}remnant}$, and remnant polarization $P_r$, respectively. Please refer to method section for cycling parameters.

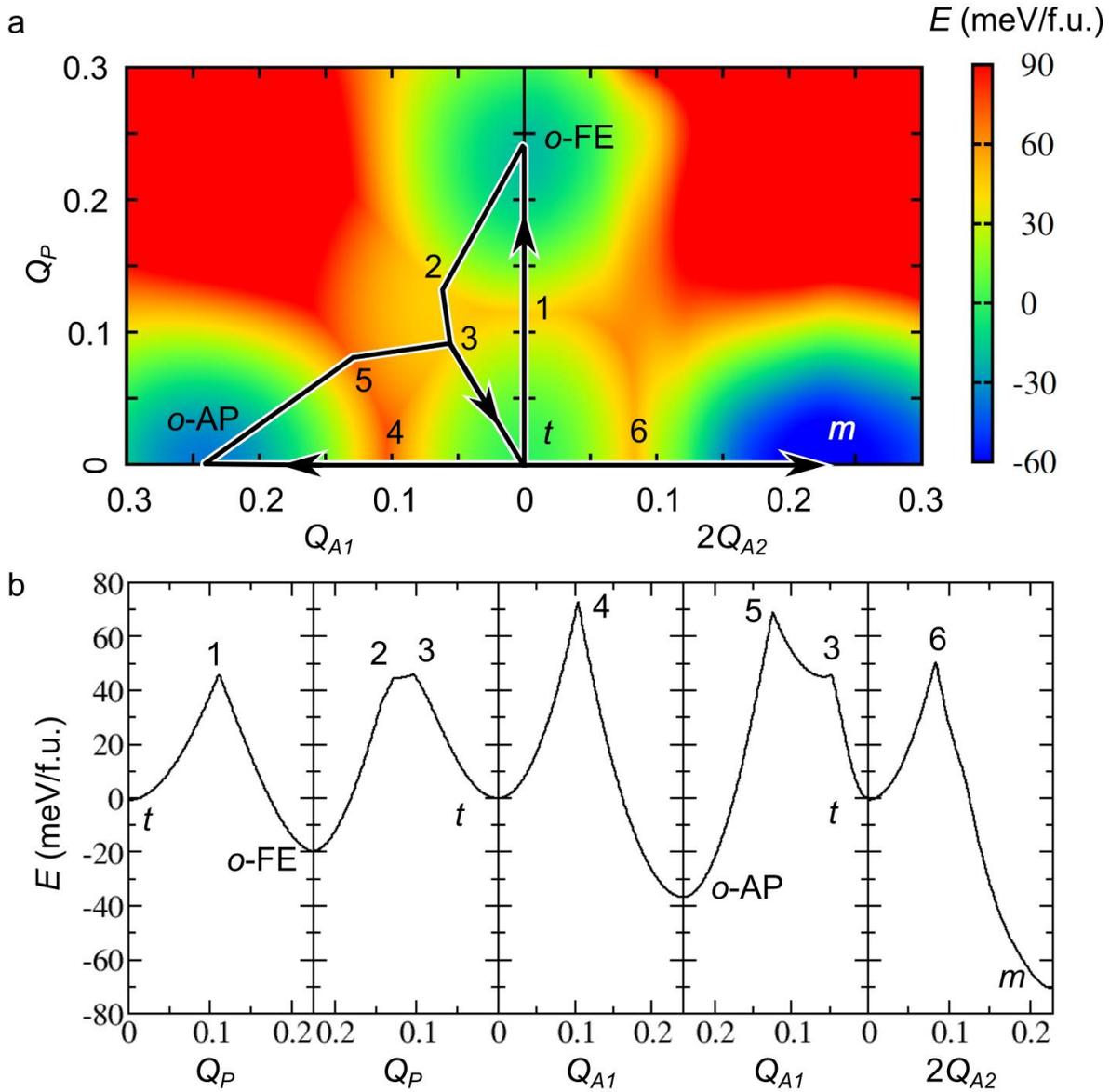

**Figure 4. Energy landscape and barriers of representative reaction paths from DFT calculations. a**, Energy landscape of bulk $HfO_2$ as a function of the amplitudes of three different lattice modes. Here, $Q_P$ is the amplitude of a polar mode, and $Q_{A1}$ and $Q_{A2}$ are the amplitudes of two different anti-polar modes. **b**, Energy profiles of several representative reaction paths linking the high-temperature tetragonal phase to the low-temperature phases.

**Table 1. The structure parameters of 12 % $HfO_2$:Y from elastic neutron diffraction performed at room temperature (290 K). Refined by 100 % $Pbc2_1$ ($\chi^2$ = 15.08 %).**

| $a$(Å) | $b$(Å) | $c$(Å) | Atom | $x$ | $y$ | $z$ | Occupancy | Uiso |
|---|---|---|---|---|---|---|---|---|
| 5.1019(3) | 5.1540(5) | 5.1036(4) | Hf | 0.0003(9) | 0.2620(1) | 0.1926(4) | 0.881(6) | 0.0143(3) |
| | | | O1 | 0.2223(2) | 0.0301(9) | 0.9231(4) | 0.997(1) | 0.0537(9) |
| | | | O2 | 0.2490(6) | 0.4928(6) | 0.4451(3) | 0.952(2) | 0.0243(8) |
| | | | Y | 0.0003(9) | 0.2620(1) | 0.1926(4) | 0.118(4) | 0.0143(3) |

**Table 2. The energy barriers $\Delta E$ of the representative reaction paths shown in Figure 4, with and without the lattice constants fixed. Please refer to Supplementary Figure 10 for the energy profiles of HfO$_2$ with lattice constants fixed. The unit of energy is meV/formula unit.**

| # of the path | 1 | 2 | 3 | 4 | 5 |
|---|---|---|---|---|---|
| Phase transitions | $t \rightarrow o$-FE | $t \rightarrow o$-FE | $t \rightarrow o$-AP | $t \rightarrow o$-AP | $t \rightarrow m$ |
| $\Delta E$ (lattices unfixed) | 46.0 | 46.0 | 72.8 | 68.8 | 50.4 |
| $\Delta E$ (lattices fixed) | 60.0 | 54.0 | 105.9 | 90.3 | 360.4 |

*Supplementary Information*

**Kinetically-stabilized Ferroelectricity in Bulk Singlecrystalline HfO$_2$:Y without Wake-up Effects**

*Xianghan Xu, Fei-Ting Huang, Yubo Qi, Sobhit Singh, Karin M. Rabe, Dimuthu Obeysekera, Junjie Yang, Ming-Wen Chu, and Sang-Wook Cheong*

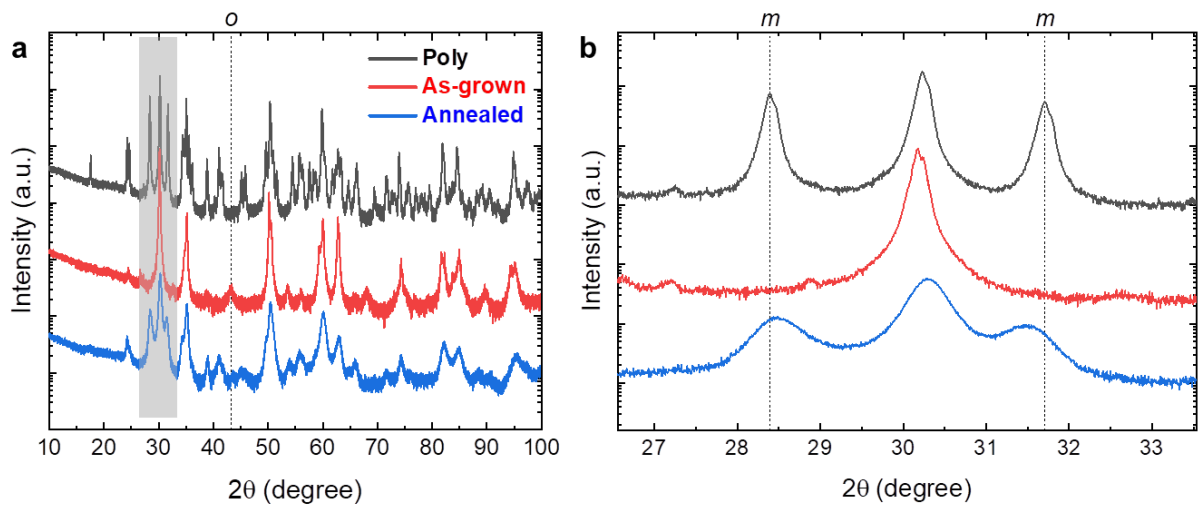

**Supplementary Figure 1. XRD of polycrystalline, as-grown and annealed 12 % HfO$_2$:Y samples. a**, The black, red and blue curve displays XRD of polycrystalline, LDFZ-growth single crystal and the crystal after 1,600°C 10 hours annealing, respectively. The vertical dashed line indicates the peak which uniquely belongs to *o*-FE phase, showing only the LDFZ-growth single crystal contains *o*-FE phase. **b**, A zoomed-in range including two monoclinic allowed reflections (11-1)$_m$ and (111)$_m$ at 2θ = 28.5° and 31.7°. As indicated by vertical dashed lines, monoclinic phase reappears after 1600°C 10 hours annealing, reflecting the meta-stability of *o*-FE phase.

.

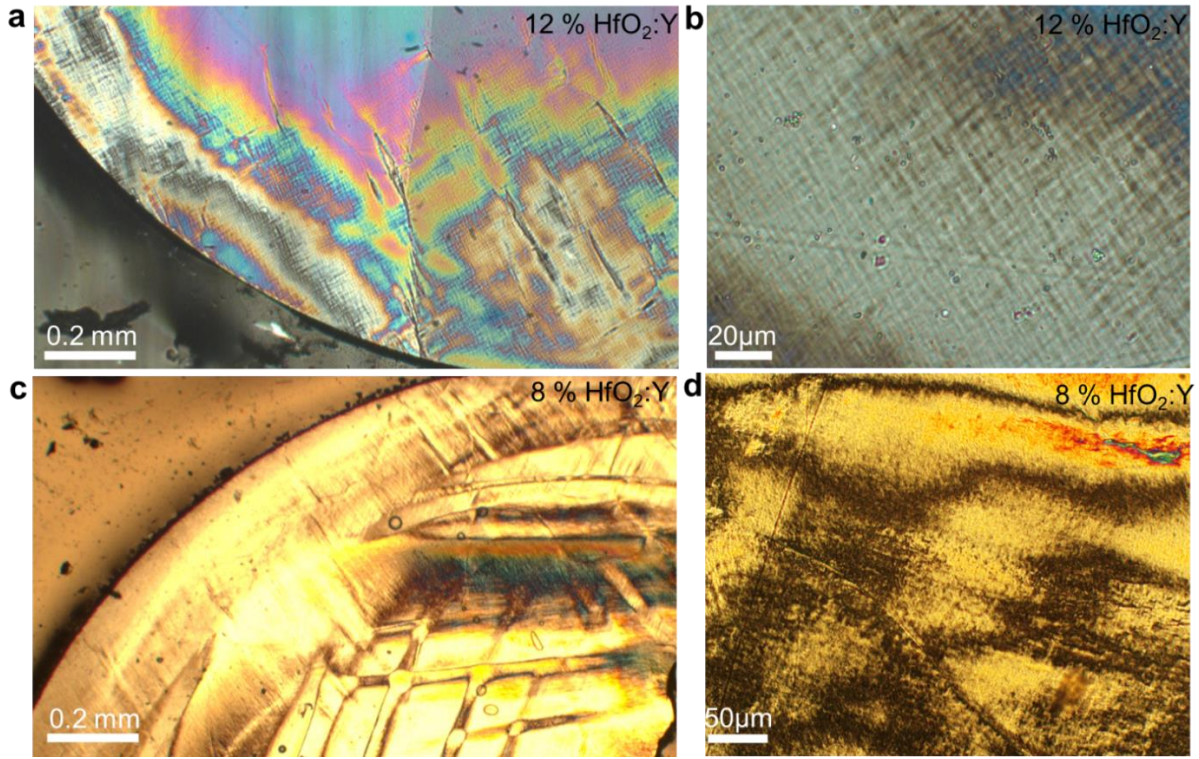

**Supplementary Figure 2. Transmission polarized optical microscope (tPOM) of HfO$_2$:Y crystals. a-b**, 12 % HfO$_2$:Y and **c-d**, 8 % HfO$_2$:Y, showing distinct crisscross of needle-like domains of one cutting surface. 12 % HfO$_2$:Y belongs to the pure *o*-FE phase while a mixed phases of cubic, monoclinic and *o*-AP phases contribute to the complicated local features in 8 % HfO$_2$:Y.

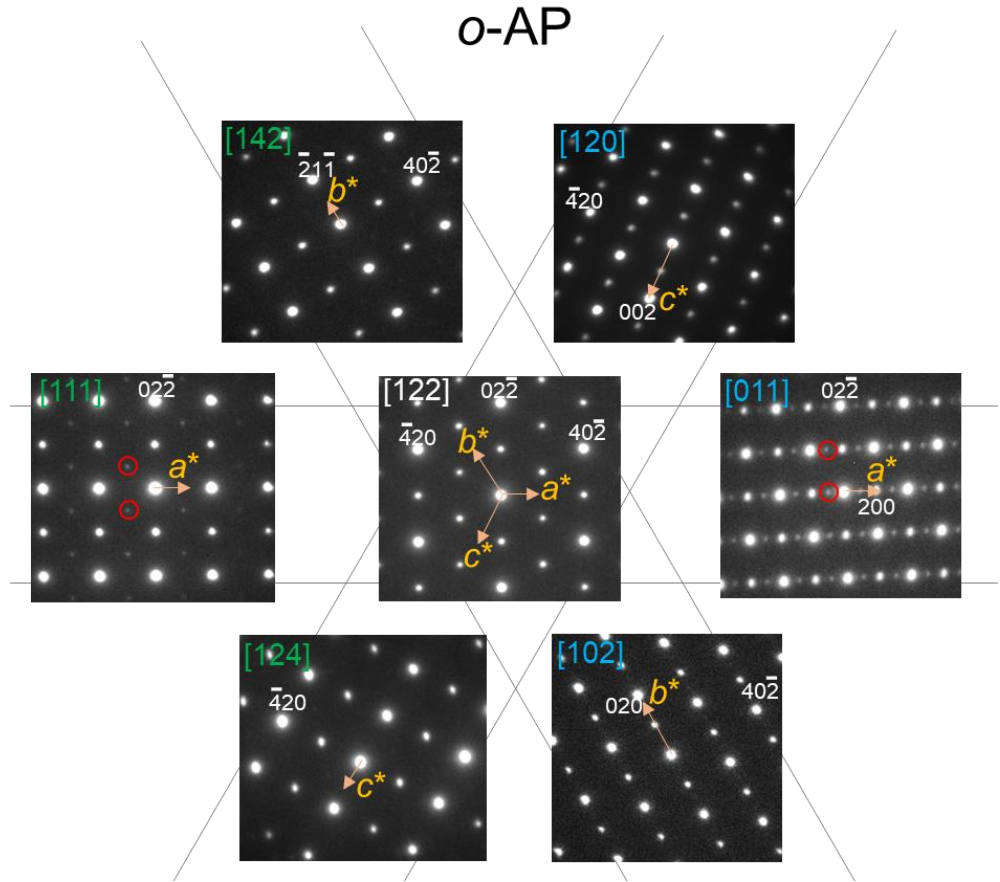

**Supplementary Figure 3. Procedure for the identification of *o*-AP phase of 8 % HfO$_2$:Y.** Beginning with the cutting surface along [122]$_{o\text{-AP}}$ = [111]$_{cubic}$, we took a series of SADPs with different zone axes. The additional peaks marked with red circles shown in [011] and [111] zone axes unambiguously indicate the cell doubling along *a*-axis, i.e., *o*-AP phase.

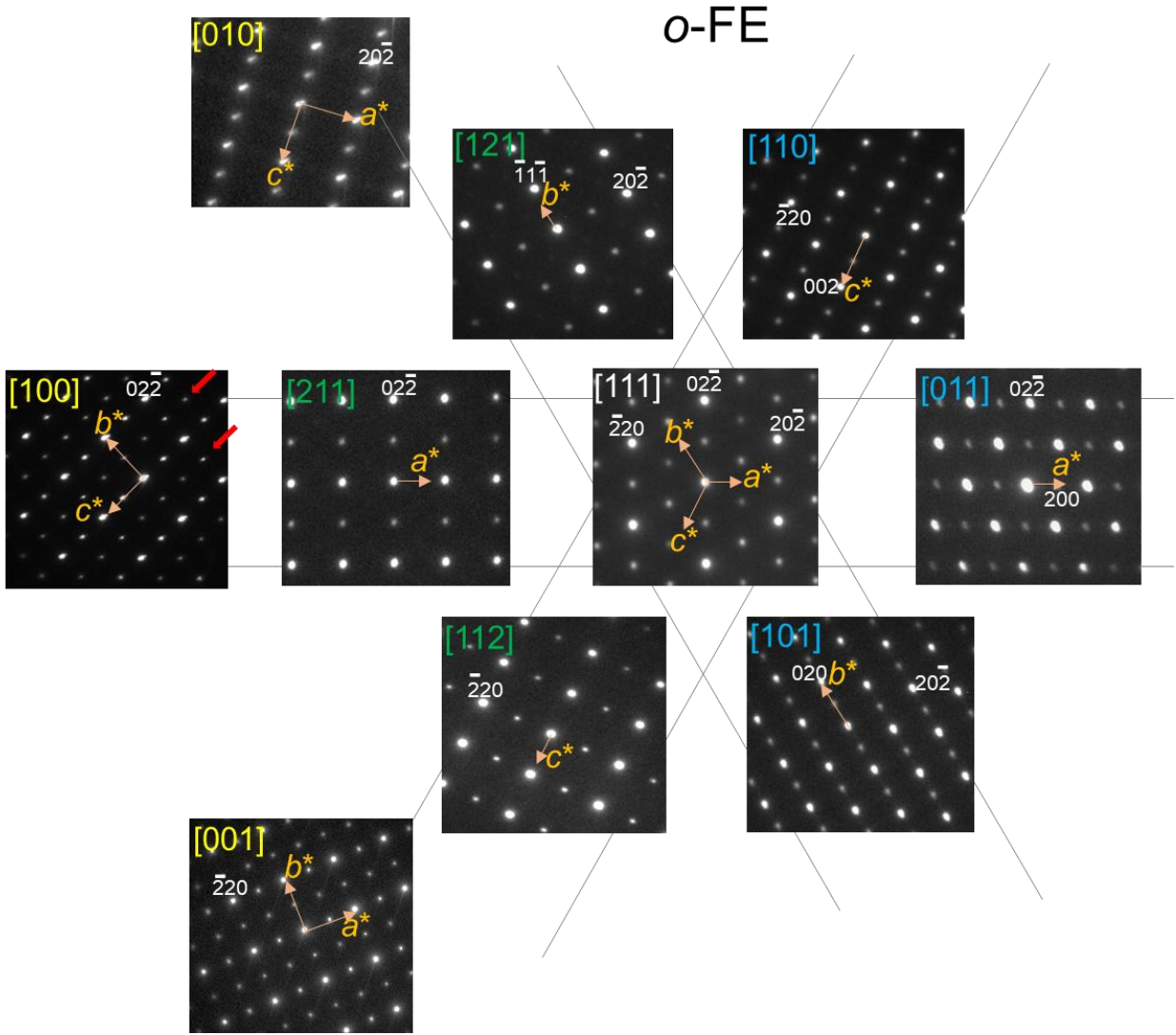

**Supplementary Figure 4. Procedure for the identification of *o*-FE phase of 12 % HfO$_2$:Y.** Four different cutting crystals were examined including [111]$_{o\text{-FE}}$, [010]$_{o\text{-FE}}$, [100]$_{o\text{-FE}}$ and [001]$_{o\text{-FE}}$ axes. Beginning with the [111]$_{o\text{-FE}}$ oriented piece, we took a series of SADPs to identify lattice paramters and symmetry. The observed orthorgonal features exclude the existence of monoclinic and tetragonal phases, independent to the x-ray data. As discussed in the main text and Supplementary Figure 5, [100]$_{o\text{-FE}}$ and [001]$_{o\text{-FE}}$ zone axes are entangled due to the existence of fine scale *ac* twins and leads to *ab*/*bc* twins along [100]$_{o\text{-FE}}$ and [001]$_{o\text{-FE}}$. For example, the additional peaks indicated by red arrows in [100]$_{o\text{-FE}}$ are attributed from the *ab* domains.

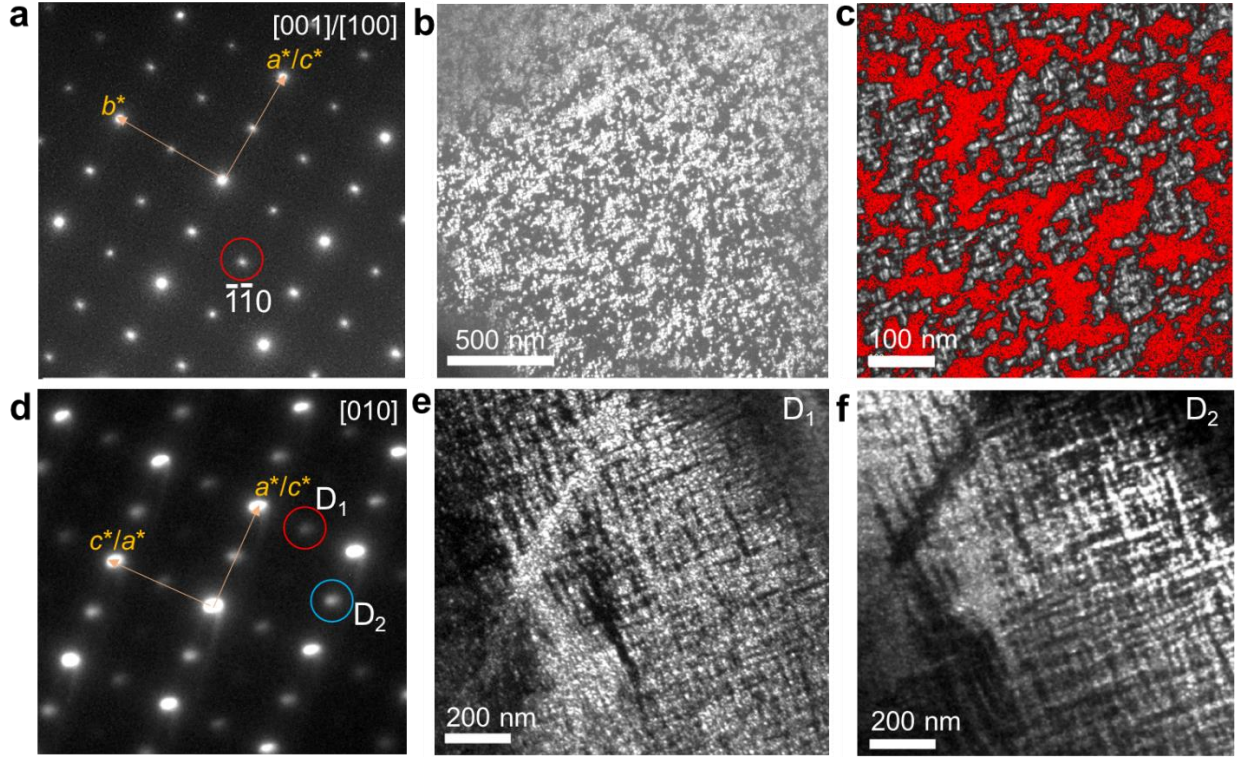

**Supplementary Figure 5. SAED and DF-TEM images of 12 % HfO$_2$:Y. a**, SAED obtained from the [001]/[100]$_{o\text{-FE}}$ zone axes. The red circle marks the Bragg peak, g = $\bar{1}\bar{1}0$, for DF-TEM imaging. **b-c**, DF-TEM images obtained using g = $\bar{1}\bar{1}0$ at different magnifications. The [001]-oriented domains, i.e. *ab* domains are map out with bright contrast since the g = $\bar{1}\bar{1}0$ is only allowed in [001] $_{o\text{-FE}}$. Areas with dark contrast correspond to *bc* domains, respectively. A false red color maps the *bc* domains in **c**. The *ab* and *bc* domains are estimated to be in half-half distibution. This is also the oriented piece that the P-E measurment was performed. **d**, SAED obtained from the [010]$_{o\text{-FE}}$ zone axe. The circled spots D$_1$ = 20$\bar{1}$ (red) and D$_2$ = $\bar{2}$01 (blue) evidently show 90$^o$ crystallographic-twin relation. **e**, DF-TEM images taken using the D$_1$ spot for imaging. **f**, DF-TEM images taken using the D$_2$ spot in the same area. Note that larger *ac* twin domains still exist in the crystal but the average sizes of *ac* twins, *ab* and *bc* domains are about few hundred nanometers square.

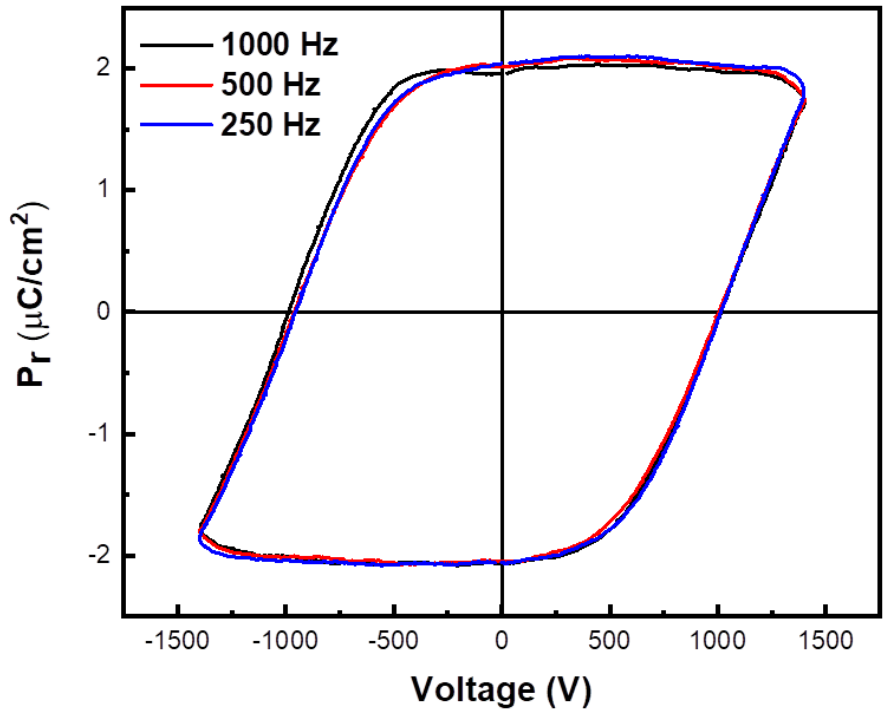

**Supplementary Figure 6. P-E loops measured with various frequencies.** The P-E loops were measured on a $[100]/[001]_{o\text{-FE}}$ polished 12 % $HfO_2$:Y crystal with thickness 5 μm by the "PUND" hysteresis loop method provided in the Ferroelectric Material Test System (RADIANT TECHNOLOGIES INC.). The black, red, and blue loops were measured with frequency 1000 Hz, 500 Hz, and 250 Hz, respectively.

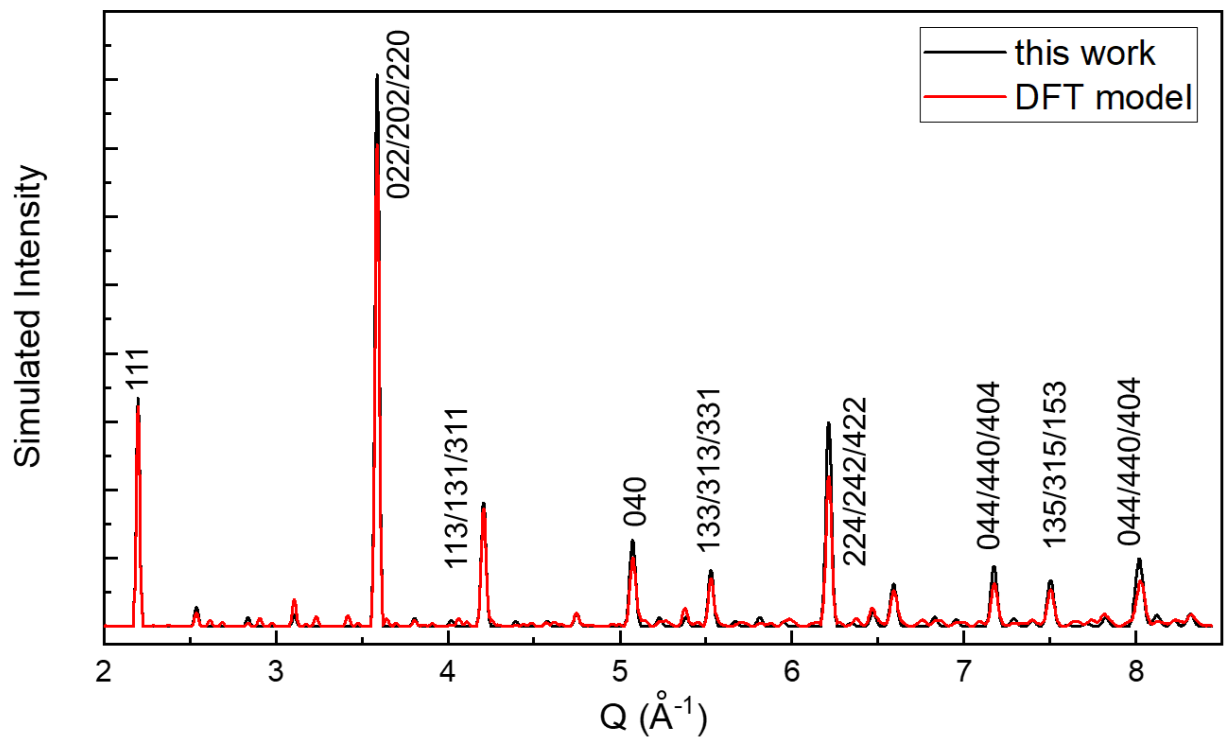

**Supplementary Figure 7. The simulated neutron pattern of the refined model and the DFT model with the reflections labelled.** The Powdercell software was utilized for the simulation. Only the atomic position and occupancy were considered for the simulation.

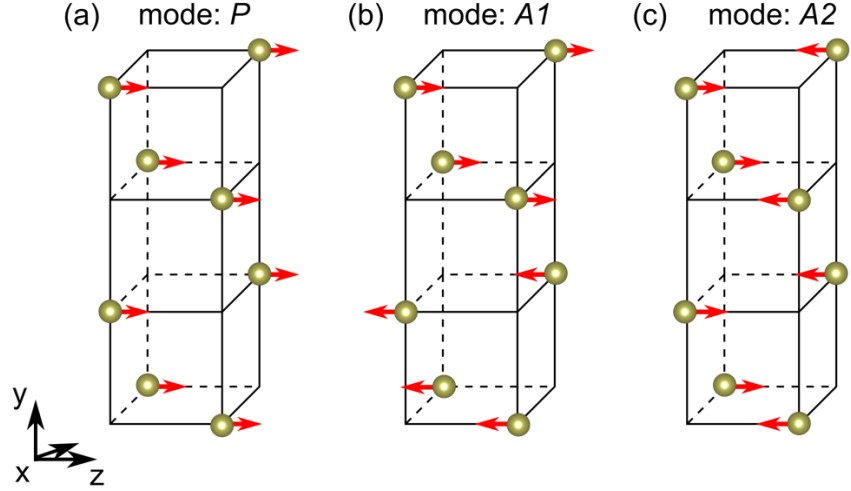

**Supplementary Figure 8. Descprition of the lattice modes.** Arrows represent the displacement of Hf atoms (olive) with respect to the center of its eight nearest neighbor oxygen atoms. In real structures, the oxygen atoms displace much more than the Hf atoms. Here, we mark the arrows on the Hf atoms for a clearer view. Here, mode $P$ is a polar mode, and modes $A1$ and $A2$ are the two anti-polar modes. The amplitude of a lattice mode is defined as $Q = \frac{1}{N}\sum_i^N d_i$, where $d_i$ is the displacement of the Hf atom $i$, and $i$ runs over all the $N$ atoms in a unit cell. $d_i$ is defined as $d_i = s(\mathbf{R}_{\text{Hf},i} - \mathbf{R}_{\text{O},i}) \cdot \hat{\mathbf{n}_i}$. Here, $\mathbf{R}_{\text{Hf},i}$ is the position of Hf atom $i$, $\mathbf{R}_{\text{O},i}$ is the position of the center of the eight nearest neihgbor oxygen atoms of Hf atom $i$, $\hat{\mathbf{n}_i}$ is a unit vector corresponding to the Hf atom $i$, whose direction is marked in the above schematic plots **a-c**, $s = 2/\sqrt{6}$ is a scaling factor.

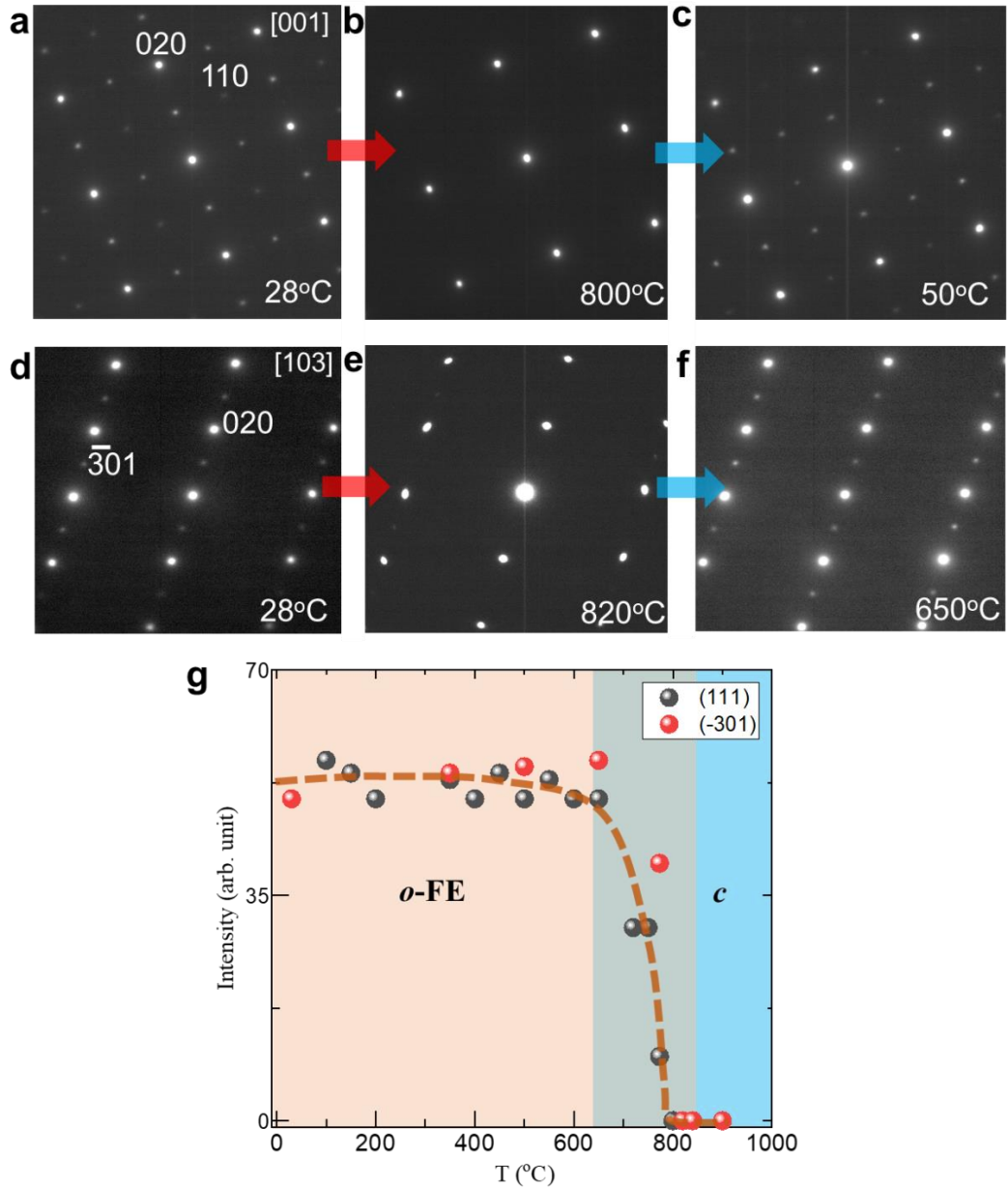

**Supplementary Figure 9. The *o*-FE to cubic structural phase transition of 12 % HfO$_2$:Y. a-c,** In-situ thermal sequence of SAED patterns of 12 % HfO$_2$:Y along [001]$_{o\text{-FE}}$, indicating a *o*-FE to cubic structural transition at ~800±50°C. The clear thermal hysteresis during warming and cooling cycles is the characteristic of first-order phase transition. **d-f,** A second round in-situ experiment was performed and tilted to [103]$_{o\text{-FE}}$ zone axis, which is 17° away from the [001]$_{o\text{-FE}}$. The SAED pattern observation along [103]$_{o\text{-FE}}$ ambigueously excludes the existence of tetragonal (*t*) phase as commonly assigned to be the high-temperature phase in doped HfO$_2$ thin films. Superlattice spots related to *o*-FE phase disappear and reappear upon warming and cooling across the transition. **g,** Intensity profiles of the g = 111 and = $\bar{3}\,0\,\bar{1}$ peaks as a function of temperature.

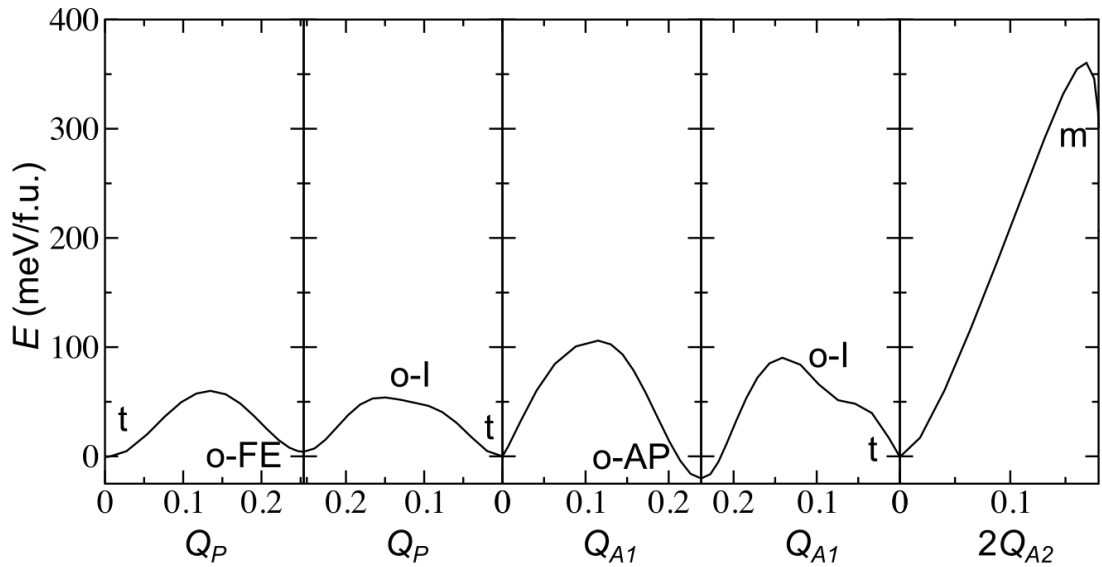

**Supplementary Figure 10. The energy profiles of HfO$_2$ with lattice constants fixed, calculated by the NEB method implemented in the Quantum-espresso package.** The energy barrier between the *t* and *m* phases increases dramatically, since the optimized *m*-phase has a shear compared to the *t* phase.

**The necessity of the kinetic theory for the stabilization of *o*-FE phase in bulk hafnia:**

As we argued in the main text, the *o*-FE phase can never be energetically stabilized (or become more favorable) over the monoclinic *m* phase simply by Y substitution and/or insertion of oxygen vacancies in bulk hafnia. Instead, a joint effort of quenching (rapid cooling) and Yttrium supersaturation is required for the kinetic stabilization of the *o*-FE phase. A rapid cooling promotes the system to overcome the lowest energy barrier, as shown in Supplementary Figure 11(a), and favors the stabilization of the *o*-FE phase. On the other hand, slow cooling gives the system enough time to overcome the highest energy barrier, and thereby, assists the stabilization of the lowest energy *m* phase in bulk.

In order to corroborate the above argument, we performed additional DFT calculation on all the possible structures of HfO$_2$:Y for 12.5% Y-concentration, and checked the energetic stability of each structure. Our calculations reveal that there is not even a single case where the *o*-FE or *o*-AP phase energetically wins over the *m*-phase.

We used the "supercell"[1] program to generate all the possible symmetry inequivalent structures of Y-substituted HfO$_2$ for 12.5 % Y-concentration while ensuring the charge neutrality condition, which requires missing of one oxygen per two Y-substitutions on Hf sites. We used the LDA optimized pristine unit cells of pure HfO$_2$ for creation of HfO$_2$:Y supercells. Then, we relax the atomic coordinates in each supercell until the residual forces on each atom are less than $10^{-3}$ eV/Å, and compute the formation energy ($E_{form}$) of each optimized structure in reference to the monoclinic ground state (*P2$_1$/c*) of pure HfO$_2$ and cubic ground state (*Pa-3*) of pure Y$_2$O$_3$ using the following expression:

$$E_{form} = E_{total}(HfO_2{:}Y) - (1-x)E_m(HfO_2) - xE_c(Y_2O_3).$$

Here, $E_{total}$(HfO$_2$:Y), $E_m$(HfO$_2$), and $E_c$(Y$_2$O$_3$) represent the total energy (eV/atom) of HfO$_2$:Y supercell, monoclinic ground state (*P2$_1$/c*) of pure HfO$_2$, and cubic ground state (*Pa-3*) of pure Y$_2$O$_3$, respectively. $x$ denotes the Y-concentration in the hafnia matrix ($x = 0.125$ in our case). Supplementary Figure 11(b) shows the relative formation energy of the five studied phases (*m*, *t*, *c*, *o*-AP, and *o*-FE) of HfO$_2$:Y. Our calculations clearly show that the *m*- phase is lower in energy compared to the *o*-AP and *o*-FE phases. This implies that only Y-substitution and/or creation of oxygen vacancies can never energetically stabilize the *o*-AP or *o*-FE phase over the *m* phase in bulk hafnia at ambient conditions (note that *t* and *c* phases are stabilized only at high temperatures). Therefore, a kinetic model, as shown in Supplementary Figure 11(a), is essential to explain the stabilization of the *o*-FE phase in bulk $HfO_2$:Y.

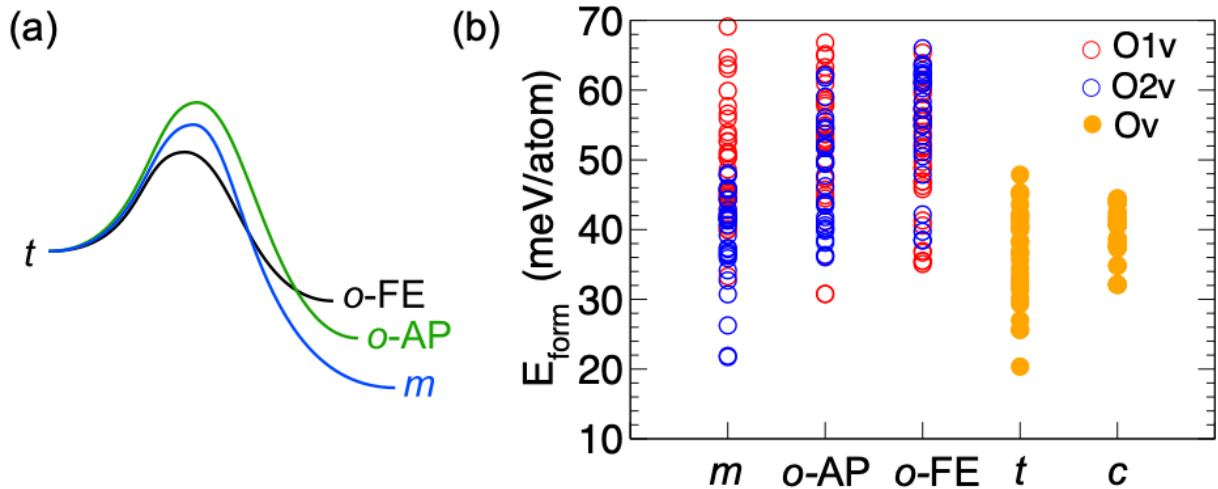

**Supplementary Figure 11. Schematic of the kinetic model and the relative formation energies of different phases of $HfO_2$:Y implying the necessity of the kinetic model for stabilization of the *o*-FE phase. a**, Schematic of the energy barrier profile of the structural phase transition from the high symmetry *t* phase to the low symmetry *m*, *o*-AP, and *o*-FE phases. The black (blue) curve indicates the favorable energy path at rapid (slow) cooling conditions. **b**, The DFT calculated formation energy of $HfO_2$:Y structures for five studied phases (*m*, *t*, *c*, *o*-AP, and *o*-FE). The formation energy ($E_{form}$) of all structures was evaluated in reference to the monoclinic ground state ($P2_1/c$) of pure $HfO_2$ and the cubic ground state ($Pa$-$3$) of pure $Y_2O_3$. The high energy structures having $E_{form}$ greater than 70 meV/atom were excluded from this figure. For *m*, *o*-AP, and *o*-FE phases there are two symmetry inequivalent oxygen sites. Red/Blue hollow circles indicate whether the oxygen vacancy was inserted at the O1/O2 site while creating the $HfO_2$:Y supercells. For *t* and *c* phases there is only one symmetry inequivalent oxygen site for the insertions of oxygen vacancies.

**Theoretical Methods:**

We use the VASP package[2,3] for calculations of total energy and forces of $HfO_2$:Y supercells generated using the "supercell" program[1]. Pristine $HfO_2$ unit cells for monoclinic *m* ($P2_1/c$), tetragonal *t* ($P4_2/nmc$), cubic *c* (*Fm-3m*), antipolar orthorhombic *o*-AP (*Pbca*), and polar orthorhombic *o*-FE ($Pca2_1$) phases were fully optimized using the LDA functional in VASP until the residual forces on each atom were less than $10^{-3}$ eV/Å. A Monkhorst-pack *k*-mesh of size $8 \times 8 \times 8$ was employed for sampling the reciprocal space of pristine hafnia unit cells, and a commensurate *k*-mesh was employed for the $HfO_2$:Y supercells. 500 eV was set as the kinetic energy cutoff for plane waves. We considered eleven valence electrons of Y ($4s^2 4p^6 5s^2 4d^1$), four valence electrons of Hf ($5d^2 6s^2$), and six valence electrons of oxygen ($2s^2 2p^4$) within the projected-augmented wave (PAW) framework[4,5].

In order to simulate the correct stoichiometry of *x*=12.5 % Y-substitution in the hafnia matrix, we used a 16 f.u. large unit cell of $HfO_2$, i.e., $Hf_{16}O_{32}$. We substituted two Hf atoms by two Y atoms and created an oxygen vacancy to maintain the charge neutrality, i.e., $Y_2Hf_{14}O_{31}$ composition. There are $^{16}C_2$ (=120) ways to distribute two Y atoms on sixteen Hf sites. Also, in the case of low symmetry hafnia structures, i.e., *m*, *o*-FE, and *o*-AP phases, there are two symmetry inequivalent oxygen sites, O1 and O2, having sixteen oxygen atoms at each site. Thus, there are $^{16}C_1$ (=16) ways to create one oxygen vacancy at each O1 and O2 sites. However, in the case of the high symmetry *t* and *c* phases, there is only one symmetry inequivalent oxygen site, hence, there are $^{32}C_1$ (=32) ways to create one oxygen vacancy at the available 32 oxygen sites. Using the "supercell" program, we generated structures considering all the aforementioned possible combinations. We also employed the crystalline symmetries to reduce the total number of generated supercell structures by merging the symmetry equivalent structures. For instance, the total number of possible structures was reduced from 1,920 to 120 using the crystal symmetries of the *o*-FE phase. In total, several hundreds symmetry inequivalent $HfO_2$:Y structures were generated for all the five studied phases. We optimized all these structures, computed their formation energies with respect to the monoclinic ground state of pure $HfO_2$ and cubic ground state of pure $Y_2O_3$, and checked the relative energetic stability of the lowest energy structures of all the five studied phases (*m*, *t*, *c*, *o*-AP, and *o*-FE) of $HfO_2$:Y (see Supplementary Figure 11).